\pdfoutput=1 % For pdflatex
\documentclass[prl,twocolumn,preprintnumbers,superscriptaddress,showpacs]{revtex4-1}

\usepackage{graphicx}% Include figure file
\usepackage{feynmp-auto} % make diagrams
\usepackage{bm}% bold math
\usepackage{amsmath}
\usepackage{lineno}
\usepackage{amssymb}
\usepackage[utf8]{inputenc}
\usepackage[colorlinks,pdfstartview=FitH]{hyperref}
\hypersetup{linkcolor=blue,citecolor=blue,filecolor=black,urlcolor=blue}

\newcommand{\be}{\begin{equation}}
\newcommand{\ee}{\end{equation}}

\begin{document}

\preprint{MIT-CTP/4884}

\author{Artur Avkhadiev}
\email{artur\_avkhadiev@brown.edu}
\affiliation{Department of Physics, Brown University, Providence RI 02912, USA}

\author{Andrey V. Sadofyev}
\email{sadofyev@mit.edu}
\affiliation{Center for Theoretical Physics, Massachusetts Institute of Technology, Cambridge, MA 02139, USA}
\affiliation{ITEP, Bolshaya Cheremushkinskaya 25, Moscow, 117218, Russia}

\title{Chiral Vortical Effect for Bosons}

\begin{abstract}
The thermal contribution to the chiral vortical effect is believed to be related to the axial anomaly in external gravitational fields. We use the universality of the spin-gravity interaction to extend this idea to a wider set of phenomena. We consider the Kubo formula at weak coupling for the spin current of a vector field and derive a novel anomalous effect caused by the medium rotation: the chiral vortical effect for bosons. The effect consists in a spin current of vector bosons along the angular velocity of the medium. We argue that it has the same anomalous nature as in the fermionic case and show that this effect provides a mechanism for helicity transfer, from flow helicity to magnetic helicity.
\end{abstract}

\maketitle

\section{Introduction}

Recently, macroscopic manifestations of the axial anomaly have attracted significant attention in the literature (for review, see \cite{Kharzeev:2012ph, Kharzeev:2015znc}). This triangle loop diagram violates classical conservation of the axial charge in the presence of electromagnetic (EM) fields (see e.g. \cite{Bell:1969ts}): 

\begin{eqnarray}
\label{anomaly}
\partial_\mu J^\mu_A=\frac{1}{2\pi^2} E\cdot B\,,
\end{eqnarray}
where $J_A$ is the axial current.

In a chiral medium, it results in vector and axial currents directed along the magnetic field or local angular velocity: chiral effects.  Moreover, the chiral vortical effect (CVE) survives in the absence of EM fields while the theory is non-anomalous in this limit, for instance

\begin{eqnarray}
\label{cve}
J^\mu_A= \left(\frac{\mu_V^2+\mu_A^2}{2\pi^2}+\frac{T^2}{6}\right)\omega^\mu\,,
\end{eqnarray}
where $\mu_{V(A)}$ is a vector (axial) chemical potential, $\omega^\mu=\frac{1}{2}\epsilon^{\mu\nu\alpha\beta}u_\nu\partial_\alpha u_\beta$ is vorticity and $u_\mu$ is the 4-velocity of the fluid element. There is an intensive discussion on the origin of  (\ref{cve}), see e.g. \cite{Vilenkin:1980zv, Son:2009tf, Sadofyev:2010is, Neiman:2010zi, Landsteiner:2011cp, Kirilin:2012sd, Braguta:2013loa, Buividovich:2013jba, Golkar:2015oxw, Flachi:2017vlp}. It is also known that this effect may result in event-by-event contributions to parity and charge parity violating observables in heavy ion collisions (see e.g. \cite{Kharzeev:2015znc}).

Despite the anomaly, it is possible to introduce a conserved generalization of the axial charge \cite{Avdoshkin:2014gpa}. Including the contribution of the anomalous effects, the axial charge is given by

\begin{eqnarray}
\label{dq5}
\partial_t \left(N_5+\mathcal{H}_{mh}+\mathcal{H}_{mfh}+\mathcal{H}_{fh}\right)=0
\end{eqnarray}
where $\mathcal{H}_{fh}=\int \left(\frac{\mu_V^2+\mu_A^2}{2\pi^2}+\frac{T^2}{6}\right)\omega^0d^3x$ is the flow helicity, $\mathcal{H}_{mh}=\frac{1}{4\pi^2}\int A\cdot Bd^3 x$  is the magnetic helicity, $\mathcal{H}_{mfh}=\frac{1}{2\pi^2}\int \mu_V v\cdot Bd^3 x$  is the mixed magnetic/flow helicity, and $J_5^0=N_5$ is the difference between the number of right and left particles. Note that the macroscopic contributions in (\ref{dq5}) are of topological origin: these helicities measure the linkage between field and flow lines. In this expression it may seem that the flow helicity $\mathcal{H}_{fh}$ can be transferred into the chiral asymmetry. On the other hand, there is no known microscopic mechanism to support such a process. Indeed, the anomaly is unmodified by medium effects (at finite temperature and density). The issue persists in the other transition of the axial charge between the macroscopic terms in (\ref{dq5}).

In this article, we concentrate on the latter issue in a neutral medium ($\mu_{V(A)}=0$), asking the following question: {\it Is it possible to find a microscopic mechanism to transfer} $\mathcal{H}_{fh}$ {\it to} $\mathcal{H}_{mh}$? If the answer is {\it``yes,''} then there is at least an indirect way to generate chiral asymmetry from a helical flow via the anomaly caused by an intermediate generation of $\mathcal{H}_{mh}$. But first, it is instructive to concentrate on the origin of the thermal contribution to (\ref{cve}), henceforth referred to as the tCVE. This effect is believed to be connected with the gravitational cousin of the axial anomaly (see e.g. \cite{AlvarezGaume:1983ig}), consisting in an axial charge non-conservation due to an external gravitational field. The coefficient in front of the gravitational anomaly is argued to be connected with the coefficient in the tCVE conductivity \cite{Landsteiner:2011cp, Landsteiner:2011iq}. Although this picture is widely accepted, note that there is an ongoing discussion on other possible origins of this effect \cite{Golkar:2012kb, Golkar:2015oxw, Kalaydzhyan:2014bfa}. In particular, there are strong arguments in favor of the relation between the tCVE and global anomalies in effective field theory \cite{Golkar:2012kb, Golkar:2015oxw}.

It is known that the gravitational anomaly is a more general phenomenon taking place for chiral bosons as well \cite{Dolgov:1987xv, Dolgov:1987yp, Dolgov:1988qx, Vainshtein:1988ww}. Moreover, due to the features of gravitational interaction, this anomaly can appear in the axial current constructed on fields of arbitrary spin. Thus, following the conjectured connection of the gravitational anomaly and the tCVE, one expects to find other anomalous effects for fields with $s\neq \frac{1}{2}$.

With this motivation, we begin our consideration of the relevant spin currents looking for effects analogous to the fermionic tCVE. We explore the novel anomalous transport via the example of the chiral vortical effect for vector bosons (bCVE), concentrating on the direct and instructive derivation relying on the Kubo formula in the weakly coupled limit. In full similarity with \cite{Landsteiner:2011cp, Golkar:2012kb}, the corresponding conductivity is expressed as

\begin{eqnarray}
\label{kubo}
\sigma_{\mathcal{V}}=\lim_{p_k\to0}\epsilon_{ijk} \frac{-i}{2p_k}\left\langle K^i T^{0j}\right\rangle|_{\omega=0}\,,
\end{eqnarray}
where $K^\mu=\epsilon^{\mu\nu\alpha\beta}A_\nu\partial_\alpha A_\beta$ is the Chern current \footnote{Note that a similar perturbative term may appear as an effective vertex in an anomalous theory \cite{Hou:2012xg, Golkar:2012kb}. Here, however, we are interested in a more general setup, particularly in the absence of fermions.}. 

As in the fermionic case, the corresponding conductivity appears as a resummation of a divergent series of 3d ``conductivities'' indicating its relation to the global anomalies \cite{Golkar:2012kb, Golkar:2015oxw}. We further propose that there is a wider set of novel anomalous effects in spin currents constructed on higher spin fields. While the arguments on their relation with the gravitational anomaly are rather convincing in the holographic setup, further field-theoretic study is required. However, it should be stressed that the bCVE and the tCVE have the same origin, and that it is evident at the level of the derivation.

When considering the bCVE for photons in a slowly varying helical motion of the medium, one finds a process generating $\mathcal{H}_{mh}$ out of $\mathcal{H}_{fh}$. Therefore, the answer for the question above is {\it``yes.''} The process of helicity transfer $\mathcal{H}_{fh}\to\mathcal{H}_{mh}$ extends the generalized axial charge picture \cite{Avdoshkin:2014gpa,Yamamoto:2015gzz,Zakharov:2016lhp} to bosonic theories, where the helical charge $\mathcal{H}_{mh}$ could have a non-electromagnetic nature. We also argue that the generation of magnetic helicity provides a microscopic mechanism to produce chiral asymmetry out of the helical motion of the medium.

This article is organized as follows. First, we discuss the gravitational anomaly in $K^\mu$ for vector bosons and its similarities with the fermionic case. We use this analogy as a guiding principle to motivate the study of polarization effects for vector bosons. Then we show that a rotating bosonic system exhibits a novel contribution to $K^\mu$ (bCVE), which is analogous to the tCVE in $J_A^\mu$. We state that the similarity between gravitational anomalies hints at a possibly deeper connection between the two chiral vortical effects and argue that they have the same origin. Finally, we use bCVE to establish a microscopic mechanism for helicity transfer. 

\section{Gravitational anomaly}
The gravitational anomaly for fermions is a consequence of the well-known triangle loop diagram in external gravitational fields \cite{AlvarezGaume:1983ig}. Once it is taken into account, the divergence of the axial current reads

\begin{eqnarray}
\label{ganf}
\nabla_\mu J_5^\mu=-\frac{1}{192\pi^2}\epsilon^{\mu\nu\alpha\beta}R^\lambda_{\rho\mu\nu}R^\rho_{\lambda\alpha\beta}
\end{eqnarray}
where $J_5^\mu=\bar\psi\gamma^\mu\gamma_5\psi$ and $\psi$ is the massless Dirac field. 

This result can be generalized to the case of other massless fields running in the triangle loop. Indeed, for vector bosons such a diagram gives (see e.g. \cite{Dolgov:1987yp, Vainshtein:1988ww})

\begin{eqnarray}
\label{ganb}
\nabla_\mu K^\mu=\frac{1}{96\pi^2}\epsilon^{\mu\nu\alpha\beta}R^\lambda_{\rho\mu\nu}R^\rho_{\lambda\alpha\beta}\,,
\end{eqnarray}
where $K^\mu=\frac{1}{\sqrt{-g}}\epsilon^{\mu\nu\alpha\beta}A_\nu\partial_\alpha A_\beta$. This anomaly is a particular example of a wider set of phenomena which is tied to the universality of the spin-gravity interaction (see e.g. \cite{AlvarezGaume:1983ig, Vainshtein:1988ww}). Note that despite the non-zero divergence $\nabla_\mu K^\mu=\frac{1}{2}F^{\mu\nu}\tilde{F}_{\mu\nu}$, chirality is conserved for massless vector bosons and $\langle \nabla_\mu K^\mu\rangle$, na\"{\i}vely, vanishes in external gravitational fields.

Some reservations should be made in the case of gauge bosons. The divergence $\partial_\mu K^\mu$ is a gauge invariant quantity, while the Chern current is not. However, a constant bulk current is always a sloppy concept: it is not the constant bulk current that is measured, but rather the change in the charge, tied to the gauge invariant divergence. In the following section, we derive perturbatively the anomalous contribution to $K^\mu$ constructed on a vector boson field. When gauge bosons are involved, we assume that the observable is the change in the charge happening, say, at the boundary of the system. Note that the charge $\int K^0d^3x$ is invariant under local gauge transformations and gives $\pm1$ for right- (left-) polarized ``photons,'' counting the difference between the number of left and right particles, while the gauge non-invariant current $K^i$ should be understood only as a contribution to the 4-divergence. One may also reformulate that in terms of the change in the angular momentum when polarized photons leave the medium.

The two anomalies (\ref{ganf}) and (\ref{ganb}) are fully analogous in the sense of spin current. It is argued in \cite{Dolgov:1987yp} that if one introduces an infinitesimal mass for fermions and vector bosons, which is always an eligible procedure in any consideration of the anomaly, then both currents are connected with relativistic generalizations of one-particle Pauli-Lubanski pseudovector. Further, we keep this limiting picture in mind when taking the massless limit in the final results.

It is worth mentioning that the origin of the tCVE is under active discussion. In the hydrodynamic limit, the gravitational anomaly involves higher derivatives giving no contribution of the first order to the axial current. Some possible solutions \cite{Jensen:2012kj} and alternative anomalous origins \cite{Golkar:2015oxw, Golkar:2012kb, Kalaydzhyan:2014bfa} are suggested in the literature. Despite this issue, there is a connection between the gravitational anomaly coefficient and the tCVE conductivity \cite{Landsteiner:2011cp} which becomes explicit in the holography \cite{Landsteiner:2011iq}. Henceforth we employ this picture, in particular, as a motivation.

\section{Kubo formula}

\begin{figure}
	\centering
	\includegraphics[width=0.3\textwidth]{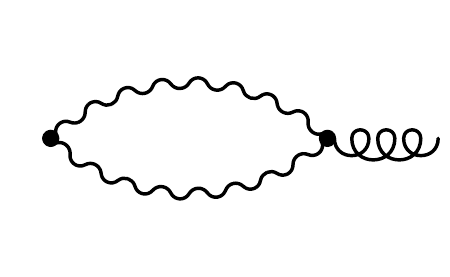}
	\caption{\label{fig1} The two-point function of the vector field spin current and the stress-energy tensor.}
\end{figure}

This section is focused on the derivation of the vortical conductivity for vector bosons in the limit of a weakly interacting medium: a gas of vector bosons. The operators in the corresponding Kubo formula (\ref{kubo}) read

\begin{eqnarray}
&~&K^\mu=\epsilon^{\mu\nu\alpha\beta}A_\nu\partial_\alpha A_\beta\notag\\
&~&T^{\mu\nu}=F^{\mu\lambda}F_{~\lambda}^{\nu}-g^{\mu\nu}\left(\frac{1}{4}F^2-\frac{1}{2}m^2A^2\right)\,.
\end{eqnarray}
We remind the reader that $\sigma_{\mathcal{V}}$ is defined by the equilibrium behavior of the corresponding retarded two-point function at zero frequency and in the small momentum limit (for details see \cite{Landsteiner:2011cp}):

\begin{eqnarray}
G_R^{i,0j}(\omega,p)|_{\omega=0}=i\epsilon_{ijk}p_k\sigma_{\mathcal{V}}+O(p^2)\,.
\end{eqnarray}
The two-point function at zero frequency is given by the Euclidean Green's function $G_R^{i,0j}=-i\mathcal{G}^{i,0j}=\left\langle K^i T^{0j}\right\rangle$. The relevant leading diagram is given in Fig.\ref{fig1}.

To avoid ambiguity, we begin by considering a massive vector field. In the calculation below, we employ the Proca formalism, taking the massless limit at the end of the procedure. According to the Matsubara technique, the two-point function reads

\begin{eqnarray}
\mathcal{G}^{\mu,0\alpha}(0,\vec p)=\beta^{-1}\sum_{n}\int\frac{d^3q}{(2\pi)^3}\frac{\epsilon^{\mu\nu\rho\sigma}p_\sigma q_\rho \left(q^\alpha\delta^0_\nu+q^0\delta_\nu^\alpha\right)}{\left((p-q)^2+m^2\right)\left(q^2+m^2\right)}\notag
\end{eqnarray}
where $p_0=0$ and $q_0$ is the bosonic thermal frequency running in the loop: $q_0=\omega_n=2\pi i T n$. For brevity, we also omit the terms that result in zero contributions on symmetry grounds. It is worth mentioning that the same diagram with photon lines is gauge independent. The issues originate in treating $K^\mu$ by itself, not this particular contribution.

For massless fermions, the effect of interest is known to be related to the $\zeta$-function resummation of conductivities in 3d Euclidean theories. This result provides an argument towards another origin of the tCVE based on global anomalies \cite{Golkar:2015oxw, Golkar:2012kb}. We find it instructive to partially generalize this statement. The two-point function above is clearly divergent in UV and it has to be regularized. The derivation is simplified in the dimensional regularization where it is sufficient to relate some momentum integrals \footnote{$\int\frac{d^3l}{(2\pi)^3}\frac{l_il^j}{(l^2-\omega^2)^2}=\omega^2\delta^j_i\int\frac{d^3l}{(2\pi)^3}\frac{1}{(l^2-\omega^2)^2}$}. Then, expanding in powers of $p$ and taking the $m\to0$ limit, one finds

\begin{eqnarray}
\mathcal{G}^{i,0j}(0,\vec p)=-\beta^{-1}\sum_{n}\epsilon^{ikl}p_l\int\frac{d^3q}{(2\pi)^3}\frac{\delta_k^j\omega_n^2+q_k q^j}{(\omega_n^2-\vec{q}^2)^2}\,.
\end{eqnarray}
Finally, in similarity with the fermionic case, the bCVE conductivity reads

\begin{eqnarray}
\sigma_{\mathcal{V}}=\frac{1}{2}T^2\sum_{n=-\infty}^\infty|n|=\frac{T^2}{12}\,.
\end{eqnarray}
Note that the factor of $\frac{1}{12}$ in the bCVE conductivity is due to the $\zeta$-function regularization of a formally divergent sum (see \cite{Golkar:2012kb}). In the case of finite mass, the result is modified and tends to zero in the limit of infinite mass-to-temperature ratio; in what follows we assume the massless limit. A similar feature is present in the case of the chiral separation effect: a magnetic-field-driven contribution to the fermionic axial current (see \cite{Metlitski:2005pr}). Finally, for the resulting spin current of massless vector bosons we have

\begin{eqnarray}
\label{bcve}
\vec{K}=\frac{T^2}{6}\vec\Omega\,,
\end{eqnarray}
where $\vec\Omega$ is the local angular velocity of the medium. We stress that the coefficient in bCVE is tied to the gravitational anomaly for bosons as much as in the fermionic case \cite{Landsteiner:2011cp, Landsteiner:2011iq, Jensen:2012kj} and that this perspective may be extended to higher spin effects.

Notably, one may arrive at the same result by following the one-point function calculation. This procedure is analogous to the axial current derivation in the fermionic case \cite{Vilenkin:1980zv}, and we omit it here.

\section{Helicity transfer}
Considering the bCVE for photons in the hydrodynamic limit, we are interested in the magnetic helicity change. As mentioned above, the Chern current is a gauge-dependent quantity requiring careful treatment. However, a gauge transformation cannot influence the divergence $\partial_\mu K^\mu$, which, in turn, defines the helicity change $\int d^3 x~\partial_\mu K^\mu\sim\partial_t\mathcal{H}_{mh}$.

One expects that multiple photons of the same polarization result in a helical ``condensate'' equivalent to a non-zero $\mathcal{H}_{mf}$ in the considered region of the space. Thus, in a na\"ive approximation, the current (\ref{bcve}) generates regions of opposite helical charges at the boundaries. This idea could be extended to a local form with the relativistic version of (\ref{bcve}), given by

\begin{eqnarray}
K^\mu=\frac{T^2}{6}\omega^\mu\,.
\end{eqnarray}
Relying on the vector boson condensation and turning on the hydrodynamic perturbations, we find

\begin{eqnarray}
\langle E\cdot B\rangle=\partial_\mu\frac{T^2}{12}\omega^\mu\,.
\end{eqnarray}
Consequently, a flow with $\frac{\partial}{\partial t}\mathcal{H}_{fh}\neq0$ is expected to generate $\mathcal{H}_{mh}$ \footnote{For discussion of chiral asymmetry generation in the presence of a temperature gradient, see \cite{Basar:2013qia, Zakharov:2016lhp}.}. 

This process is of particular interest since it provides a mechanism to generate microscopic asymmetry through a change in macroscopic motion of the medium via an intermediate generation of $\langle E\cdot B\rangle$. It is instructive to note the connection of the discussion above with the tCVE renormalization (see \cite{Hou:2012xg, Golkar:2012kb}). There, it is pointed out that in the presence of dynamical gauge fields, the tCVE reads

\begin{eqnarray}
\label{tcvecor}
J_5^\mu=\frac{T^2}{6}\left(1+\frac{e^2}{4\pi^2}\right)\omega^\mu\,.
\end{eqnarray}
The coupling $e$ is explicitly restored to make the relation of the second term to the axial anomaly evident. This result supports the proposed mechanism of helicity transfer to chirality and provides an example of an interesting interplay between the bosonic anomalous spin current and its fermionic counterpart. Indeed, the condensate of polarized photons results in an additional anomalous contribution to the axial current. Note that the corresponding coefficient is fixed by two multiples coming from different anomalous phenomena. Other chiral effects may also be modified, see discussions in \cite{Kirilin:2012mw, Khaidukov:2013sja, Kirilin:2013fqa, Jensen:2013vta, Gorbar:2013upa, Kalaydzhyan:2014bfa}.

In the theory of magnetohydrodynamics (MHD), it is well known that the dynamics of a magnetic fluid considerably depends on topological properties of the field and flow configuration. Particularly, in the ideal MHD limit, magnetic field lines are frozen in the medium volume element and the magnetic helicity is conserved. In other words, the electric field is completely screened and there is no way to change $\mathcal{H}_{mh}$. When the ideal limit of infinite conductivity is relaxed, the helicity is changing due to reconnections. However, this process is slower than magnetic energy dissipation and the helicity constrains the dynamics of the system. The bCVE current connects the two helicities in a plasma of light bosons and may result in a considerable modification of the MHD evolution.

The photon properties are modified in the medium, including the Debye mass screening by the Coulomb interaction. In the case of gluons or other non-Abelian vectors, this issue is additionally complicated by a possible generation of a magnetic mass. These effects depend on the coupling and disappear in the non-interacting limit. In this note we restrict ourselves to the leading order contributions and leave a detailed analysis of possible modifications of the conductivity by the interaction for future considerations. Note, however, that the connection of chiral effects with the anomaly results in an additional robustness, which is also supported in the strongly coupled limit by holographic considerations. In the case of the bCVE, one may rely on a bottom-up construction for a strongly interacting theory with no fermions. The key properties of such a model involving the relevant gravitational anomaly remain the same as in \cite{Landsteiner:2011iq}, and the axial (spin) current gains the bCVE contribution connected with its anomaly.

\section{Discussions}

In this note, we propose a set of novel anomalous effects consisting in spin currents of various fields along vorticity. We argue that these effects have a common origin for different spin values and can be thought of as a generalization of the tCVE. We explicitly derive the anomalous contribution for the case of bCVE in the weakly coupled limit using the Kubo formula.

We further discuss the connection between the bCVE and the corresponding anomaly (\ref{ganb}). While the relation of the tCVE with the gravitational and global anomalies is under discussion, the arguments based on the holographic picture \cite{Landsteiner:2011iq} are rather convincing. The gravitational anomaly takes place for fields of any spin (see e.g. \cite{AlvarezGaume:1983ig, Vainshtein:1988ww}). We argue that the universality of spin-gravity interaction provides a strong argument in favor of the existence of other chiral effects in the spin currents for $s>1$ fields. 

Inverting the chain of arguments, we see no contradiction in the relation between the tCVE and the gravitational anomaly. Indeed, comparing the derivations of the tCVE and the bCVE one concludes that these effects have a common origin which is expected to be the same for higher spins as well. The similarity of effects in spin currents is a feature of the gravitational interaction, in agreement with the picture above. It would be interesting to study $s>1$ contributions in detail to probe the underlying anomalous dynamics (see e.g. \cite{Golkar:2015oxw, Son:2009tf, Landsteiner:2011cp}) -- especially the relation of these anomalous effects to the global anomalies \footnote{Methods of \cite{Golkar:2015oxw} were extended to chiral scalars in $d=2$ and gravitinos in $d=2, 6$, see \cite{Chowdhury:2015pba}}.

We show that the bCVE provides a mechanism to produce magnetic helicity out of the medium helical motion. In turn, the axial anomaly caused by polarized photons can generate chiral asymmetry, supporting (\ref{dq5}). That extends the idea of the correction to the tCVE by gauge bosons (\ref{tcvecor}), which may be thought of as an interplay between the bosonic and fermionic vortical effects. The bCVE current also mixes flow and magnetic helicities, resulting in a more involved and rich dynamics. For MHD in chiral plasma, where microscopic and macroscopic helicities play a crucial role, the interplay between them is a matter of principle.  Note that this relation between the helicities also takes place in systems with no fermions. 

We emphasize that various systems may exhibit the bCVE and higher spin anomalous effects. For instance, the bCVE could take place in P-odd condensed matter systems such as P-wave superconductors and superfluids, QGP, and in cosmological primordial plasma. In particular, it would be interesting to study the effect on gluons at high temperatures and their contribution to QGP polarization effects. It should also be mentioned that anomalous transport can take place in certain cold atom systems with spin-orbit coupling (see e.g. \cite{Huang:2015mga}). These systems may also exhibit the bCVE. Finally, we stress that it would be interesting to study the bCVE and higher spin effects in the chiral kinetic theory \cite{Stephanov:2012ki, Son:2012zy, Mueller:2017arw}. {\it Note added --} Soon after we released this work, the first study of the bCVE in the context of chiral kinetic theory appeared in \cite{Yamamoto:2017uul}.

\section{Acknowledgments}
Authors are grateful to V. Kirilin, D. Kharzeev, K. Rajagopal and Y. Yin for useful discussions. AS thanks V. Zakharov for pointing out the possibility of bosonic anomalies which resulted in the idea of this work. The work of AS was supported by the U.S. Department of Energy under grant Contract No. DE-SC0011090. The work on sections II and III is supported by Russian Science Foundation Grant No 16-12-10059.

\bibliography{limit_v}

\end{document}